\documentstyle[11pt,a4]{article}

\newcommand{\be}{\begin{equation}}
\newcommand{\ee}{\end{equation}}
\newcommand{\bea}{\begin{eqnarray}}
\newcommand{\eea}{\end{eqnarray}}
\newcommand{\bed}{\begin{displaymath}}
\newcommand{\eed}{\end{displaymath}}

\newcommand{\bc}{\begin{center}}
\newcommand{\ec}{\end{center}}

\newcommand{\br}{({\bf r})}
\newcommand{\brp}{({\bf r'})}
\newcommand{\text}{\rm}
\begin{document}

\begin{sf}
\thispagestyle{empty}
\bc
{\tt $\bullet$To appear in International Journal of Quantum 
Chemistry$\bullet$ }
\ec

\vspace{0.5cm}

\begin{quote}

\begin{flushleft}
 
{\bf \LARGE \sf The Optimized Effective Potential Method of Density Functional
Theory: Applications to Atomic and Molecular Systems}

\vspace{1.5cm}

{\Large \sf T.~Grabo and E.K.U.~Gross}

\vspace{0.5cm}

Institut f\"ur Theoretische Physik, Universit\"at W\"urzburg,
Am Hubland, D-97074 W\"urzburg, Germany

\vspace{1cm}

{\bf \sf Abstract}

Using the optimized effective potential method
in conjunction with the semi-analytical approximation due to Krieger,
Li and Iafrate, we have performed fully self-consistent
exact exchange-only density-\-functional calculations for diatomic
molecules with a fully numerical basis-set-free molecular code.
The results
are very similar to  the ones obtained with the Hartree Fock approach.
Furthermore we present results for ground states of positive
atomic ions including correlation contributions in the approximation
of Colle and Salvetti. It is found
that the scheme performs significantly better than
conventional Kohn-Sham calculations.

\end{flushleft}
\end{quote}
\vspace{1cm}


\section{\sf  Introduction}

Since its development by Talman and Shadwick \cite{TalmanShadwick:76},
following the original  
idea of Sharp and Horton \cite{SharpHorton:53}, the optimized effective
potential (OEP) 
method has been recognized 
\cite{SahniGruenebaumPerdew:82,PerdewNorman:82}
as the exact implementation of
exchange-only density functional theory (DFT) 
\cite{HohenbergKohn:64, KohnSham:65, DreizlerGross:90,
ParrYang:89}. Due to the rather large   
computational effort involved, this scheme has not been used
extensively. In the last years, however, the situation has changed. Using an
accurate analytical approximation due to Krieger, Li and Iafrate (KLI)
\cite{KriegerLiIafrate:92, KriegerLiIafrate:92a, LiKriegerIafrate:93} the 
effort involved in numerical calculations based on the OEP has become 
comparable
to conventional Kohn-Sham calculations, while the gain in
accuracy is considerable. 

In the following, we will briefly review the theoretical foundations of the
OEP method and the KLI approximation and present, in section
\ref{Exchange-only results for molecular systems}, 
applications of the method to molecular systems. Finally, in
section \ref{Correlation contributions to the OEP}, calculations for
atomic systems including correlation effects are discussed and
compared with conventional Kohn-Sham results.

\section{\sf  Basic formalism} \label{Basic formalism}

We start from ordinary spin DFT \cite{BarthHedin:72,PantRajagopal:72}, 
where the basic variables are the spin-up and spin-down densities
$\rho_{\uparrow} \br$ and $\rho_{\downarrow}\br$, respectively. They are
obtained by self-consistently solving the single-particle
Schr\"odinger equations (atomic units are used throughout)
\be \label{1p-eq}
\left( -\frac{{\bf \nabla}^2}{2} + 
       V_{\sigma}[\rho_{\uparrow},\rho_{\downarrow}] \br \right)
   \varphi_{j \sigma} \br 
= 
\varepsilon_{j \sigma} \varphi_{j \sigma} \br 
\qquad 
j = 1, \ldots, N_{\sigma} 
\qquad 
\sigma = \uparrow, \downarrow 
\ee
where
\be
\rho_{\sigma}({\bf r})
=
\sum_{i=1}^{N_{\sigma}} \vert \varphi_{i \sigma}({\bf r}) \vert^{2}, 
\ee
The Kohn-Sham potentials $V_\sigma \br$ may be written in the usual way
as
\be \label{vtot}
V_{\sigma}\br = v_{\rm ext}\br + \int d^3 r' \ 
 \frac{ \rho({\bf r'}) }{ \vert {\bf r} - {\bf r'} \vert }
  + V_{{\rm xc}\sigma}\br ,
\ee
\be
\rho \br 
=
\sum_{\sigma = \uparrow, \downarrow} \rho_{\sigma}\br
\ee
where $v_{\rm ext} \br$ represents the Coulomb potential of the nuclei
and $V_{{\rm xc}\sigma}\br $ is a {\em local\/} exchange-correlation (xc)
potential formally defined as functional derivative of the xc energy
\be \label{vxc}
 V_{{\rm xc}\sigma}\br :=
 \frac{ \delta E_{\rm xc}\left[\rho_{\uparrow},\rho_{\downarrow} \right]
}{ \delta \rho_{\sigma}({\bf r}) }  .
\ee
In order to understand the nature of the OEP method we
recall that the Hohenberg-Kohn theorem \cite{HohenbergKohn:64}, applied
to non-interacting systems, ensures that the ground-state determinant
and hence all occupied orbitals are unique
functionals of the spin densities:
\be \label{phi_func_den}
\varphi_{j \sigma}\br
=
\varphi_{j \sigma} [\rho_{\uparrow},\rho_{\downarrow}] \br .
\ee
As a consequence of (\ref{phi_func_den}), {\em any orbital functional\/}
$ E_{\rm xc} [ \{ \varphi_{j \tau} \}] $ {\em is an implicit functional of\/}
$\rho_{\uparrow}$ {\em and\/} $ \rho_{\downarrow}$, provided that the
orbitals come from a {\em local\/} potential. 

The starting point of the so-called OEP method is the total energy
functional 
\begin{eqnarray} \label{energie}
E_{{\rm tot}}^{\rm OEP}\left[\rho_{\uparrow},\rho_{\downarrow} \right] 
& = &
 \sum_{\sigma = \uparrow, \downarrow} \sum_{i=1}^{N_{\sigma}} 
 \int d^3 r \
\varphi_{i \sigma}^{\ast}({\bf r})
 \left(-\frac{1}{2} {\bf\nabla}^{2} \right) \varphi_{i \sigma}({\bf r})
 \nonumber \\
& & + \int d^3 r \ v_{\rm ext} \br \rho \br
\nonumber \\ 
& & +  \frac{1}{2}  \int d^3 r \int \ d^3 r' \ \frac{\rho({\bf r})
       \rho({\bf r'})}{\vert {\bf r - r' \vert}} 
        \nonumber \\
& & + E_{{\rm xc}}^{\rm OEP} \left[ \{\varphi_{j \tau} \} \right]
\end{eqnarray}
where, in contrast to ordinary spin DFT, the xc energy is an {\em
explicit\/} (approximate) functional of spin orbitals and therefore 
only an {\em
implicit\/} functional of the spin densities $\rho_{\uparrow}$ and
$\rho_{\downarrow}$. As a consequence of this fact, the calculation of
the xc potentials from Eq. (\ref{vxc}) is somewhat more
complicated: We use the chain rule for functional derivatives to obtain
\bea
 V_{{\rm xc}\sigma}^{\rm OEP} \br 
& = &
 \frac{ \delta E_{\rm xc}^{\rm OEP}\left[\{\varphi_{j \tau} \}  \right]
      }{ \delta \rho_{\sigma}({\bf r}) } \nonumber \\
& = &  \sum_{\alpha = \uparrow, \downarrow} \sum_{i = 1}^{N_{\alpha}} 
      \int d^3 r' \
    \frac{ \delta E_{\rm xc}^{\rm OEP} \left[ \{ \varphi_{j \tau} \} \right]
      }{ \delta \varphi_{i\alpha}({\bf r'}) }
    \frac{ \delta \varphi_{i\alpha}({\bf r'}) 
      }{ \delta \rho_{\sigma}({\bf r) } }
   + c.c. 
\eea
and, by applying the functional chain rule once more,
\be \label{vxc_double_chain}
 V_{{\rm xc}\sigma}^{\rm OEP} \br 
=
 \sum_{\alpha = \uparrow, \downarrow} \sum_{\beta = \uparrow, \downarrow} 
  \sum_{i = 1}^{N_{\alpha}} \int d^3 r'  \int d^3 r'' \
   \left(
    \frac{ \delta E_{\rm xc}^{\rm OEP} \left[ \{ \varphi_{j\tau} \} \right]
      }{ \delta \varphi_{i\alpha}({\bf r'}) }
    \frac{ \delta \varphi_{i\alpha}({\bf r'}) 
      }{ \delta V_{\beta}({\bf r''}) } + c.c. 
   \right) \frac{ \delta V_{\beta}({\bf r''}) 
      }{ \delta \rho_{\sigma}({\bf r}) } .
\ee
The last term on the right-hand side is readily identified with the
inverse $\chi_{\rm s}^{-1} \left( {\bf r},{\bf r'} \right)$ of the
density response function of a system of non-interacting particles
\be \label{chi_def}
\chi_{\rm s \alpha, \beta} \left( {\bf r},{\bf r'} \right)
:=
  \frac{ \delta \rho_{\alpha}\br }{ \delta V_{\beta} \brp}.
\ee
This quantity is diagonal with respect to the spin variables so that
Eq. (\ref{vxc_double_chain}) reduces to 
\be \label{vxc_der_chi}
 V_{{\rm xc}\sigma}^{\rm OEP} \br 
=
 \sum_{\alpha = \uparrow, \downarrow} 
 \sum_{i = 1}^{N_{\alpha}} \int d^3 r'  \int d^3 r'' \
   \left(
    \frac{ \delta E_{\rm xc}^{\rm OEP} \left[ \{ \varphi_{j\tau} \} \right]
      }{ \delta \varphi_{i\alpha}({\bf r'}) }
    \frac{ \delta \varphi_{i\alpha}({\bf r'}) 
      }{ \delta V_{\sigma}({\bf r''}) } 
     + c.c. 
   \right) \chi_{\rm s \sigma}^{-1} \left( {\bf r''},{\bf r} \right) . 
\ee
Acting with the response operator (\ref{chi_def}) on both sides of
Eq. (\ref{vxc_der_chi}) one obtains 
\be \label{pre_oep_int_eq}
\int d^3 r' \ V_{\rm xc \sigma}^{\rm OEP} \brp 
    \chi_{\rm s \sigma} \left( {\bf r'},{\bf r} \right) 
=
 \sum_{\alpha = \uparrow, \downarrow} 
 \sum_{i = 1}^{N_{\alpha}} \int d^3 r'  \
    \frac{ \delta E_{\rm xc}^{\rm OEP} \left[ \{ \varphi_{j\tau} \} \right]
      }{ \delta \varphi_{i\alpha}({\bf r'}) }
    \frac{ \delta \varphi_{i\alpha}({\bf r'}) 
      }{ \delta V_{\sigma}({\bf r}) } + c.c. .
\ee
Finally, the second functional derivative on the right-hand side of
Eq. (\ref{pre_oep_int_eq}) is calculated using first-order
perturbation theory. This yields 
\be
\frac{\delta \varphi_{i\alpha}({\bf r'})}{\delta V_{\sigma} ({\bf r})}
=
\delta_{\alpha, \sigma}
\sum_{\stackrel{k=1}{k \neq i}}^{\infty}
\frac{ \varphi_{k\sigma}({\bf r'}) \varphi_{k\sigma}^{\ast}({\bf r})
 }{ \varepsilon_{i\sigma} - \varepsilon_{k\sigma} } 
\varphi_{i\sigma}({\bf r}) .
\ee
Using the fact that the Kohn-Sham response function can be written as
\be
\chi_{\rm s \sigma} \left( {\bf r},{\bf r'} \right) 	
=
\sum_{i=1}^{N_\sigma} \sum_{\stackrel{k=1}{k \neq i}}^{\infty}
\frac{ \varphi_{i\sigma}^{\ast}({\bf r}) \varphi_{k\sigma}({\bf r}) 
       \varphi_{k\sigma}^{\ast}({\bf r'}) 
       \varphi_{i\sigma}({\bf r'}) 
     }{
       \varepsilon_{i\sigma} - \varepsilon_{k\sigma} } + c.c.
\ee
the integral equation (\ref{pre_oep_int_eq}) takes the form
\begin{equation} \label{OEP-int}
\sum_{i=1}^{N_\sigma}
\int  d^3 r' 
\left( V_{{\rm xc} \sigma}^{\rm OEP} ({\bf r'}) - 
       u_{{\rm xc} i \sigma}({\bf r'})       \right)
G_{{\rm s} i \sigma} \left( {\bf r'},{\bf r} \right) 
\varphi_{i\sigma} \br \varphi_{i\sigma}^{\ast} \brp	
 + c.c. = 0
\end{equation}
where
\begin{equation} \label{uxc}
u_{{\rm xc} i \sigma}({\bf r})
:=
 \frac{ 1 }{ \varphi_{i\sigma}^{\ast} ({\bf r}) }
 \frac{ \delta E_{{\rm xc}}^{\rm OEP}\left[ \{ \varphi_{j\tau} \} \right]
}{ \delta \varphi_{i\sigma}({\bf r}) }
\end{equation}
and
\be \label{KS-Green's-function}
G_{{\rm s} i \sigma} \left( {\bf r},{\bf r'} \right) 
 :=
\sum_{\stackrel{k=1}{k \neq i}}^{\infty}
\frac{ \varphi_{k\sigma}({\bf r}) \varphi_{k\sigma}^{\ast}({\bf r'})
    }{ \varepsilon_{i\sigma} - \varepsilon_{k\sigma} } .
\ee
The derivation of the OEP integral equation (\ref{OEP-int}) described
here was first given by G\"orling and Levy \cite{GorlingLevy:94}.
It is important to note that the same expression results 
\cite{TalmanShadwick:76,KriegerLiIafrate:92,GraboGross:95}
if one demands that
the local one-particle potential  appearing in Eq.
(\ref{1p-eq}) be the {\em optimized\/} one yielding orbitals minimizing the
total energy functional (\ref{energie}), i.e. that
\be \label{statcond}
\left. \frac{ \delta E_{tot}^{OEP} }{ \delta V_{\sigma}\br }
\right|_{V=V^{OEP}} = 0 .
\ee

The main advantage of the OEP method is that it allows for the {\em
exact\/} treatment of the exchange energy.
Splitting up the total xc-functional into an exchange and a correlation 
part
\be
E_{\rm xc}^{\rm OEP} \left[ \{\varphi_{j \tau} \}  \right] 
=
E_{\rm x} \left[ \{\varphi_{j \tau} \}  \right] +
E_{\rm c} \left[ \{\varphi_{j \tau} \}  \right] 
\ee
we can use the exact Fock expression
\be \label{exhf}
E_{{\rm x}}\left[ \{\varphi_{j \tau} \}  \right] =
- \frac{1}{2}\sum_{\sigma = \uparrow, 
\downarrow} \sum_{i,k=1}^{N_{\sigma}} 
\int d^3 r \int d^3 r' \  
 \frac{\varphi_{i\sigma}^{\ast}\br \varphi_{k\sigma}^{\ast}\brp
\varphi_{k\sigma}\br 
\varphi_{i\sigma}\brp}{\vert {\bf r - r'} \vert} .
\ee
Performing the functional derivative with respect to the orbitals one
obtaines for the x-part $u_{{\rm x} i \sigma} \br$ of the function 
$u_{{\rm xc} i \sigma} \br$:
\be \label{uxi-xonly}
u_{{\rm x} i \sigma} \br 
= 
- \frac{1}{\varphi_{i\sigma}^{\ast}\br} \sum_{k=1}^{N_{\sigma}}
\varphi_{k\sigma}^{\ast}\br 
\int d^3 r' 
\ \frac{\varphi_{i\sigma}^{\ast}\brp \varphi_{k\sigma}\brp
}{\vert {\bf r - r'} \vert}.
\ee

The use of the exact exchange energy has several advantages over the
conventional {\em explicitly} density dependent xc functionals. 
Most importantly it ensures
the correct $-1/r$ decay of the xc-potential for large $r$, 
reflecting
the fact that it is self-interaction free.
One has to emphasize that the OEP has the correct $-1/r$ tail for {\em
all } orbitals, i.e. for both the occupied and the unoccupied ones. By
contrast,  the conventional Hartree-Fock (HF)
approach, which uses the same expression (\ref{exhf}) for the
exchange energy but a {\em nonlocal} potential defined via the equation
\be \label{HF-potential}
\left( \hat V_{{\rm x}\sigma}^{\rm HF} \varphi_{i\sigma} \right) \br 
= 
-\sum_{j=1}^{N_{\sigma}} \int d^3 r'
 \frac{\varphi_{j\sigma}^{\ast}\brp \varphi_{i\sigma}\brp}{\vert
{\bf r - r'} \vert}  \varphi_{j\sigma}\br
\ee 
is self-interaction free only for the {\em occupied\/}
orbitals. However, x-only OEP calculations \cite{TalmanShadwick:76,
NormanKoelling:84, WangEtAl:90, EngelChevaryMacdonaldVosko:92, 
EngelVosko:93, KriegerLiIafrate:92}, i.e. with the approximation 
$E_{{\rm c}}\left[ \{\varphi_{j \tau} \} \right] = 0$, performed on
atomic systems have shown that the results for various {\em physical}
quantities of interest such as total ground-state energies are
very similar to HF results despite the different nature of the corresponding
x-potentials. As - by construction - the HF scheme gives  the
variationally best, i.e. lowest, total energy, the x-only OEP solutions are
always slightly higher in energy. 

The solution of the full integral equation (\ref{OEP-int}) 
is numerically very demanding and has been achieved so far 
only for systems with spherical symmetry 
\cite{TalmanShadwick:76, NormanKoelling:84, WangEtAl:90,
EngelChevaryMacdonaldVosko:92, EngelVosko:93, KriegerLiIafrate:92}.
Therefore, one has to resort to further approximations for practical reasons.
Krieger, Li and Iafrate \cite{KriegerLiIafrate:92} have suggested the
analytical approximation 
\be \label{KLI-Green's-function}
G_{{\rm s}i\sigma} \left( {\bf r}, {\bf r'} \right) 
\approx 
\frac{1}{\triangle
\epsilon} \left( \delta \left( {\bf r} - {\bf r'} \right) -
\varphi_{i\sigma}\br \varphi_{i\sigma}^{\ast}\brp \right)
\ee
for the Green's-function-type quantity (\ref{KS-Green's-function}). 
Substituting this into the integral
equation (\ref{OEP-int}) and performing some algebra one arrives at
the approximate equation
\be \label{kli-eq}
 V_{{\rm xc}\sigma}^{\rm KLI} \br 
 =
 \frac{ 1 }{ 2 \rho_{\sigma}({\bf r}) }
 \sum_{i=1}^{N_{\sigma}}  \vert \varphi_{i \sigma}({\bf r})
\vert^{2}  
 \left[ u_{{\rm xc}i\sigma}({\bf r}) 
        + \left(\bar{V}_{{\rm xc}i\sigma}^{\rm KLI}  -
        \bar{u}_{{\rm xc}i\sigma}  \right)  + c.c. \right]
\ee
where $\bar u_{{\rm xc}j\sigma}$ denotes the average value of
$u_{{\rm xc}j\sigma}\br$ taken over the density of the $j\sigma$ orbital, i.e.
\be
\bar u_{{\rm xc}j\sigma} = \int d^3 r \ \vert \varphi_{j
\sigma}({\bf r}) \vert^{2}
u_{{\rm xc}j\sigma}\br 
\ee
and similarly for $\bar V_{{\rm xc}\sigma}^{\rm KLI}$.
In contrast to the exact integral equation (\ref{OEP-int}) the KLI
equation (\ref{kli-eq}) can be solved explicitly for
$V_{{\rm xc}\sigma}$ by multiplication with $\rho_{i\sigma} \br$ and
subsequent integration. This leads to linear $(N_{\sigma}-1)
\times (N_{\sigma}-1)$ equations for the unknown constants
$\left( \bar{V}_{{\rm xc}i\sigma}  - \bar{u}_{{\rm xc}i\sigma} \right)$:
\be \label{lin-eq}
\sum_{i=1}^{N_{\sigma}-1} \left( \delta_{ji} - M_{ji\sigma} \right) \left( 
\bar V_{{\rm xc}i\sigma}^{\rm KLI} - 
\frac{1}{2} \left( \bar u_{{\rm xc}i\sigma} +
\bar u_{{\rm xc}i\sigma}^{\ast} \right)  \right) 
=
\bar V_{{\rm xc}j\sigma}^{S} - 
\frac{1}{2} \left( \bar u_{{\rm xc}j\sigma} +
 \bar u_{{\rm xc}j\sigma}^{\ast} \right)
\ee
with $j= 1, \ldots, N_{\sigma}-1$,
\be
M_{ji\sigma} := \int d^3 r \  \frac{
\vert \varphi_{j\sigma}({\bf r}) \vert^{2}
\vert \varphi_{i\sigma}({\bf r}) \vert^{2}
}{\rho_{\sigma}\br} 
\ee
and
\be
\bar V_{{\rm xc}j\sigma}^{S}\br := \int d^3 r \
\frac{\vert \varphi_{j\sigma}({\bf r}) \vert^{2}}{ \rho_{\sigma}\br} 
\sum_{i=1}^{N_{\sigma}} 
\vert \varphi_{i\sigma}({\bf r}) \vert^{2} \frac{1}{2}  
\left( u_{{\rm xc}i\sigma}\br+ u_{{\rm xc}i\sigma}^{\ast}\br \right) . 
\ee
The orbitals corresponding to the highest single-particle
energy eigenvalues $\varepsilon_{N\sigma}$ have to be excluded from the linear
equations (\ref{lin-eq}) in order to ensure the correct long-range
behaviour of $  V_{{\rm xc}\sigma}^{\rm OEP}\br$
\cite{KriegerLiIafrate:92}. It is
an important property of 
the KLI approximation that it is exact for two-particle systems,
where one has only one electron per spin projection. In this case, the OEP
integral equation (\ref{OEP-int}) may be solved exactly to yield
(\ref{kli-eq}). 
Furthermore, for these systems the OEP is also identical with the HF
potential (\ref{HF-potential}).

At first sight, the KLI approximation (\ref{KLI-Green's-function}) might
appear rather crude. The final result (\ref{kli-eq}) for the KLI
potential, however, can also be understood \cite{KriegerLiIafrate:95} as a
well-defined mean-field approximation.
Explicit calculations on atoms performed in the x-only limit
\cite{KriegerLiIafrate:92, KriegerLiIafrate:92a, LiKriegerIafrate:93} 
show that the KLI-approximation yields excellent results which
differ only by a few ppm from the much more time-consuming
exact solutions of the full integral equation (\ref{OEP-int}).

\section{\sf  Exchange-only results for molecular systems}
\label{Exchange-only results for molecular systems} 

In order to demonstrate the validity of the KLI-approach for more
complex systems, we have performed x-only OEP calculations for diatomic 
molecules in KLI
approximation employing the exact exchange energy functional as
defined by equation (\ref{exhf}) and neglecting correlation effects.
 This approach will, in the following,
be referred to as {\em x-only KLI}.
Our calculations have been performed with a fully numerical
basis-set-free code, 
developed from the X$\alpha$ program written  
by Laaksonen, Sundholm and Pyykk\"o
\cite{LaaksonenPyykkoSundholm:83a, LaaksonenPyykkoSundholm:83b,
LaaksonenSundholmPyykko:85}. The code solves 
the one-particle Schr\"odinger equation for diatomic
molecules 
\be \label{1p-eq.mol}
\left( 
-\frac{{\bf \nabla}^2}{2} 
- \frac{Z_1}{\vert {\bf R}_1 - {\bf r} \vert} 
- \frac{Z_2}{\vert {\bf R}_2 - {\bf r} \vert} 
+ V_{{\rm H}} \br
+ V_{{\rm x}\sigma}^{\rm KLI} \br       
\right) 
\varphi_{j \sigma} \br 
= 
\varepsilon_{j \sigma} \varphi_{j \sigma} \br,
\ee
where $\bf R_i$ denotes the location and $Z_i$ the nuclear charge of
the $i$-th nucleus in the molecule.
The partial differential equation is solved in
prolate spheroidal coordinates on a two-dimensional mesh by a 
relaxation method, while the third variable, the
azimuthal angle, is treated analytically. The Hartree potential 
\be \label{vhart}
 V_{{\rm H}} \br = \int d^3 r' \ 
 \frac{ \rho({\bf r'}) }{ \vert {\bf r} - {\bf r'} \vert }
\ee
and the functions $u_{{\rm x} i \sigma} \br$ (cf.~Eq.~(\ref{uxi-xonly}))
needed for the calculation of the exchange potential $V_{{\rm
x}\sigma}^{\rm KLI} \br$ (cf.~Eq.~(\ref{kli-eq}))
are computed as solutions of a Poisson and
Poisson-like equation, respectively. In this step, the same relaxation
technique as for the solution of the Schr\"odinger equation
(\ref{1p-eq.mol}) is employed.
Starting with an initial guess
for the wave functions $\varphi_{i\sigma} \br$, equations
(\ref{1p-eq.mol}), (\ref{vhart}), (\ref{uxi-xonly}) together with
(\ref{kli-eq}) are iterated until self-consistent. A very detailed
description of the code is given in \cite{LaaksonenPyykkoSundholm:86}.  

In order to test the accuracy of the program, we have performed
calculations on the Beryllium and Neon atom which are compared in Table
\ref{moltest1} to exact results obtained with a one-dimensional atomic code.
The results agree to all
decimals given. Furthermore, from Table \ref{moltest2} it is evident
that our program gives results for
the two-electron molecules He$_2$ and HeH$^+$ which are identical to the
HF ones obtained by Laaksonen et 
al \cite {LaaksonenPyykkoSundholm:83b} as they should be, as the HF and 
x-only KLI schemes are identical for these systems.

\begin{table}[bthp]
\begin{center}
\begin{tabular}{|l|r|r|r|r|}
\hline
& \multicolumn{2}{c|}{Beryllium} & \multicolumn{2}{c|}{Neon} \\
\hline
& \multicolumn{1}{c|}{1D} & \multicolumn{1}{c|}{2D} & 
  \multicolumn{1}{c|}{1D} & \multicolumn{1}{c|}{2D}  \\ 
\hline
Grid & 400 & $153 \times 249$ & 400 & $153 \times 249$ \\
\hline
E$_{\rm TOT}$          & -14.5723 & -14.5723 & -128.5448 & -128.5448 \\
$\varepsilon_{\rm HOMO}$ & -0.3089  & -0.3089 & -0.8494 & -0.8494 \\
$< 1/r >$          &  2.1039  &  2.1039 & 3.1100 & 3.1100 \\
$< r^2 >$          &  4.3255  &  4.3255 & 0.9367 & 0.9367 \\
\hline
\end{tabular}
\end{center}
\caption{  X-only KLI results for the Beryllium and Neon atom. 
1D denotes exact values obtained with our
one-dimensional code,
2D the results from our two-dimensional code with one nuclear
charge set equal to zero. All 
numbers in atomic units. } 
\label{moltest1}   
\end{table}

\begin{table}[bthp]
\begin{center}
\begin{tabular}{|l|r|r|r|r|}
\hline
& \multicolumn{2}{c|}{H$_2$} & \multicolumn{2}{c|}{HeH$^+$} \\
\hline
R & \multicolumn{2}{c|}{1.4} & \multicolumn{2}{c|}{1.455} \\
\hline
& \multicolumn{1}{c|}{HF} & x-only KLI & \multicolumn{1}{c|}{HF} &
x-only KLI \\ 
\hline
E$_{\rm TOT}$           & -1.133630 & -1.133630 & -2.933103 &
-2.933103 \\ 
$\varepsilon_{1\sigma}$ & -0.594659 & -0.594659 & -1.637451 &
-1.637451 \\  
Q$_1^e$                &  0        &  0        & -0.494460 &
-0.494460 \\ 
Q$_2^e$                &  0.243289 &  0.243289 &  0.373727 &
0.373727 \\ 
Q$_3^e$                &  0        &  0        & -0.231525 &
-0.231525 \\ 
Q$_4^e$                &  0.090721 &  0.090721 &  0.173962 &
0.173962 \\ 
$< r^2 >$               &  2.573930 &  2.573930 &  1.340832 &
1.340832 \\ 
\hline
\end{tabular}
\end{center}
\caption{  Comparison of results for two electron molecules with
assumed bond length R. HF
values from
\protect\cite{LaaksonenPyykkoSundholm:83b}. Q$_1^e$, Q$_2^e$, Q$_3^e$,
Q$_4^e$,  denote the 
electronic contributions to the dipole, quadrupole, octopole and
hexadecapole moments, respectively, determined from the 
molecular midpoint. All numbers in atomic units. } 
\label{moltest2}   
\end{table}

\begin{table}[bthp]
\begin{center}
\begin{tabular}{|l|r|r|r|r|}
\hline
& \multicolumn{1}{c|}{HF} & \multicolumn{1}{c|}{x-only KLI} &
\multicolumn{1}{c|}{Slater} & \multicolumn{1}{c|}{x-only LDA}  \\ 
\hline
E$_{\rm TOT}$          & -7.9874 & -7.9868  & -7.9811  & -7.7043 \\
$\varepsilon_{1\sigma}$& -2.4452 & -2.0786  & -2.3977  & -1.7786 \\
$\varepsilon_{2\sigma}$& -0.3017 & -0.3011  & -0.3150  & -0.1284 \\
Q$_1^e$                & 0.6531  &  0.6440  &  0.8614  &  0.8679 \\
Q$_2^e$                & 7.1282  &  7.1365  &  6.9657  &  6.7717 \\
Q$_3^e$                & 2.9096  &  2.9293  &  3.0799  &  2.6924 \\
Q$_4^e$                & 16.0276 &  16.1311 &  15.5881 &  15.0789 \\
\hline
\end{tabular}
\end{center}
\caption{  X-only results for LiH. HF values for bond length of
3.015 a.u.
from  \protect\cite{LaaksonenPyykkoSundholm:86}. 
Present calculations performed on a $153 \times
193 $ grid with bond length of 3.015 a.u. All numbers in atomic units. } 
\label{LiH}   
\end{table}

\begin{table}[bthp]
\begin{center}
\begin{tabular}{|l|r|r|r|r|}
\hline
& \multicolumn{1}{c|}{HF} & \multicolumn{1}{c|}{x-only KLI} &
\multicolumn{1}{c|}{Slater} & \multicolumn{1}{c|}{x-only LDA}  \\ 
\hline
E$_{\rm TOT}$          & -25.1316 & -25.1290 & -25.1072 & -24.6299 \\
$\varepsilon_{1\sigma}$& -7.6863  & -6.8624  & -7.4837  & -6.4715 \\
$\varepsilon_{2\sigma}$& -0.6482  & -0.5856  & -0.6358  & -0.3956 \\
$\varepsilon_{3\sigma}$& -0.3484  & -0.3462  & -0.3721  & -0.1626 \\
Q$_1^e$                &  5.3525  &  5.3498  &  5.2991  &  5.3154 \\
Q$_2^e$                &  12.1862 &  12.1416 & 11.4720  & 11.9542 \\
Q$_3^e$                &  15.6411 &  15.5618 & 14.3328  & 14.0904 \\ 
Q$_4^e$                &  25.8492 &  25.4188 & 25.2152  & 21.9134 \\
\hline
\end{tabular}
\end{center}
\caption{  X-only results for BH. HF values for bond length of
2.336 a.u.
from  \protect\cite{LaaksonenPyykkoSundholm:86}. Present calculations
performed on a $193 \times 
265 $ grid with bond length of 2.336 a.u. All numbers in atomic units. } 
\label{BH}   
\end{table}

\begin{table}[bthp]
\begin{center}
\begin{tabular}{|l|r|r|r|r|}
\hline
& \multicolumn{1}{c|}{HF} & \multicolumn{1}{c|}{x-only KLI} &
\multicolumn{1}{c|}{Slater} & \multicolumn{1}{c|}{x-only LDA}  \\ 
\hline
E$_{\rm TOT}$          & -100.0708 & -100.0675 & -100.0225 & -99.1512 \\
$\varepsilon_{1\sigma}$& -26.2946 & -24.5116  & -25.6625  & -24.0209 \\
$\varepsilon_{2\sigma}$& -1.6010  & -1.3994  & -1.4327  & -1.0448 \\
$\varepsilon_{3\sigma}$& -0.7682  & -0.7772  & -0.8167  & -0.4483 \\
$\varepsilon_{1\pi}$   & -0.6504  & -0.6453  & -0.6897  & -0.3109 \\
Q$_1^{\rm tot}$        & -0.7561  & -0.8217  & -0.8502  & -0.6962 \\
Q$_2^{\rm tot}$        &  1.7321  &  1.8013  &  1.8472  &  1.7124 \\
Q$_3^{\rm tot}$        & -2.5924  & -2.7222  & -2.8782  & -2.4662 \\ 
Q$_4^{\rm tot}$        &  5.0188  &  5.1825  &  5.3720  &  4.7068 \\
$<1/r>_{\rm H}$        &  6.1130  &  6.0878  &  6.0909  &  6.0901 \\
$<1/r>_{\rm F}$        &  27.1682 &  27.1622 & 27.6049  & 27.0289 \\
\hline
\end{tabular}
\end{center}
\caption{  X-only results for FH. HF values for bond length of
1.7328 a.u.
from  \protect\cite{LaaksonenPyykkoSundholm:86}. Present calculations
performed on a $161 \times 
321 $ grid with bond length of 1.7328 a.u. All numbers in atomic units. } 
\label{HF}   
\end{table}

\begin{table}[bthp]
\begin{center}
\begin{tabular}{|l|r|r|r|r|}
\hline
& \multicolumn{1}{c|}{HF} & \multicolumn{1}{c|}{x-only KLI} &
\multicolumn{1}{c|}{Slater} & \multicolumn{1}{c|}{x-only LDA}  \\ 
\hline
E$_{\rm TOT}$          & -5.72333 & -5.72332 & -5.72332 & -5.44740 \\
$\varepsilon_{1\sigma \rm g}$& -0.92017 & -0.91929 & -0.91977 &
-0.51970 \\
$\varepsilon_{1\sigma \rm u}$& -0.91570 & -0.91566 & -0.91614 &
-0.51452 \\
Q$_2^e$                &  31.36165 & 31.35931 & 31.35907 & 31.35507 \\
Q$_4^e$                &  245.8779 & 245.8643 & 245.8615 & 245.8255 \\ 
\hline
\end{tabular}
\end{center}
\caption{  X-only results for He$_2$. HF values for bond length of 5.6 a.u.
from  \protect\cite{LaaksonenPyykkoSundholm:86}. Present calculations
performed on a $209 \times 
225 $ grid with bond length of 5.6 a.u. All numbers in atomic units. } 
\label{He2}   
\end{table}

\begin{table}[bthp]
\begin{center}
\begin{tabular}{|l|r|r|r|r|}
\hline
& \multicolumn{1}{c|}{HF} & \multicolumn{1}{c|}{x-only KLI} &
\multicolumn{1}{c|}{Slater} & \multicolumn{1}{c|}{x-only LDA}  \\ 
\hline
E$_{\rm TOT}$          & -14.8716 & -14.8706 & -14.8544 & -14.3970 \\
$\varepsilon_{1\sigma \rm g}$& -2.4531 & -2.0276 & -2.3875 & -1.7869
\\ 
$\varepsilon_{1\sigma \rm u}$& -2.4528 & -2.0272 & -2.3873 & -1.7864
\\ 
$\varepsilon_{2\sigma \rm g}$& -0.1820 & -0.1813 & -0.1989 & -0.0922
\\ 
Q$_2^e$                &  27.6362 & 27.4993 & 29.0014 & 29.4401 \\
Q$_4^e$                &  159.9924 & 159.6809 & 169.1300 & 172.8505 \\
\hline
\end{tabular}
\end{center}
\caption{  X-only results for Li$_2$. HF values for bond length of
5.051 a.u.
from  \protect\cite{LaaksonenPyykkoSundholm:86}. Present calculations
performed on a $209 \times 
225 $ grid with bond length of 5.051 a.u. All numbers in atomic units. } 
\label{Li2}   
\end{table}

\begin{table}[bthp]
\begin{center}
\begin{tabular}{|l|r|r|r|r|}
\hline
& \multicolumn{1}{c|}{HF} & \multicolumn{1}{c|}{x-only KLI} &
\multicolumn{1}{c|}{Slater} & \multicolumn{1}{c|}{x-only LDA}  \\ 
\hline
E$_{\rm TOT}$          & -29.1337 & -29.1274 & -29.0939 & -28.4612 \\
$\varepsilon_{1\sigma \rm g}$& -4.73150 & -4.09876 & -4.60353 &
-3.78576 \\
$\varepsilon_{1\sigma \rm u}$& -4.73147 & -4.09872 & -4.60351 &
-3.87571 \\
$\varepsilon_{2\sigma \rm g}$& -0.39727 & -0.33452 & -0.37659 &
-0.23067 \\
$\varepsilon_{2\sigma \rm u}$& -0.24209 & -0.23489 & -0.26524 &
-0.13163 \\
Q$_2^e$                &  46.0878 & 46.2475 & 43.4833 & 46.2501 \\
Q$_4^e$                &  261.774 & 277.365 & 249.135 & 281.950 \\
\hline
\end{tabular}
\end{center}
\caption{  X-only results for Be$_2$. HF values for bond length of
4.6 a.u.
from  \protect\cite{LaaksonenPyykkoSundholm:86}. Present calculations
performed on a $209 \times 225 $ grid with bond length of
4.6 a.u. All numbers in atomic units. } 
\label{Be2}   
\end{table}

\begin{table}[bthp]
\begin{center}
\begin{tabular}{|l|r|r|r|r|}
\hline
& \multicolumn{1}{c|}{HF} & \multicolumn{1}{c|}{x-only KLI} &
\multicolumn{1}{c|}{Slater} & \multicolumn{1}{c|}{x-only LDA}  \\ 
\hline
E$_{\rm TOT}$          & -108.9936 & -108.9856 & -108.9110 & -107.7560 \\
$\varepsilon_{1\sigma \rm g}$& -15.6822 & -14.3722 & -15.2692 &
-13.8950 \\
$\varepsilon_{1\sigma \rm u}$& -15.6787 & -14.3709 & -15.2682 &
-13.8936 \\
$\varepsilon_{2\sigma \rm g}$& -1.4726 & -1.3076 & -1.3316 &
-0.9875 \\
$\varepsilon_{2\sigma \rm u}$& -0.7784 & -0.7453 & -0.7473 &
-0.4434 \\
$\varepsilon_{3\sigma \rm g}$& -0.6347 & -0.6305 & -0.6521 &
-0.3335 \\
$\varepsilon_{1\pi \rm g}$& -0.6152 & -0.6818 & -0.6960 &
-0.3887\\
Q$_2^{\rm tot}$  &  -0.9372 & -0.9488 & -1.1756 & -1.1643 \\
Q$_4^{\rm tot}$  &  -7.3978 & -6.7476 & -7.1266 & -6.2553 \\
$<1/r>_{\rm N}$  &  21.6543 & 21.6439 & 21.9749 & 21.5820 \\
\hline
\end{tabular}
\end{center}
\caption{  X-only results for N$_2$. HF values for bond length of
2.07 a.u.
from  \protect\cite{LaaksonenPyykkoSundholm:86}. Present calculations
performed on a $209 \times 225 $ grid with bond length of
2.07 a.u. All numbers in atomic units. } 
\label{N2}   
\end{table}

For comparison, we have performed additional x-only calculations with
two other approximations of $V_{x \sigma} \br$ and $E_{x}$,
respectively. The first one of these, denoted  by
{\em Slater} in the following, uses - like the HF and x-only KLI
method - the exact orbital representation of $E_x$ given in equation
(\ref{exhf}) but the averaged exchange potential due to Slater
\cite{Slater:51} given by
\be \label{slatpot}
V_{x \sigma}^{S} \br = - \frac{1}{\rho_{\sigma}\br}
\sum_{i,j=1}^{N_{\sigma}} 
\varphi_{j\sigma}^{\ast}\br \varphi_{i\sigma} \br 
\int d^3 r' 
\ \frac{\varphi_{i\sigma}^{\ast}\brp \varphi_{j\sigma}\brp
}{\vert {\bf r - r'} \vert},
\ee
which may be obtained from (\ref{kli-eq}) by setting the constants 
$ \bar{V}_{{\rm xc}i\sigma}  - \bar{u}_{{\rm xc}i\sigma}$ equal to
zero for all $i$. The 
other is the well known x-only local density approximation (LDA) of
conventional DFT. As for the KLI calculations, we have successfully
tested our implementations on atomic systems. 

Results are given in Tables \ref{LiH} through \ref{N2} for LiH,
BH, FH, He$_{2}$, Li$_{2}$, Be$_{2}$ and N$_2$. For each system we
show the total ground state energy E$_{\rm TOT}$, the various orbital
energy eigenvalues $\varepsilon$ and the nonzero electronic
contributions to the dipole, quadrupole,
octopole and hexadecapole moments denoted by Q$_1^e$, Q$_2^e$, Q$_3^e$
and Q$_4^e$ calculated from the geometrical center of the molecule,
respectively, except for FH and N$_2$, where the total moments
(including nuclear contributions)
calculated from the center of mass are
given, denoted by Q$_L^{\rm tot}$. For these two molecules we also
present the expectation values of $1/r$, denoted by $<1/r>$,
calculated at the nuclei.

For all physical quantities of interest, i.e.~for E$_{\rm TOT}$, the
energies $\varepsilon_{\rm HOMO}$ of the highest occupied orbitals and
the multipole moments, the x-only KLI
and HF results differ only slightly, usually by a few hundredths of a
percent for total energies, a few tenths of a percent for
$\varepsilon_{\rm HOMO}$ and a few percent for the multipole
moments. The largest difference between the $\varepsilon_{\rm HOMO}$
values occurs for Be$_2$, where they differ by 3\%. For N$_2$, the
energetic order of  
the  $1\pi_{\rm u}$ and $3\sigma_{\rm g}$ orbital is reversed in all DFT
approaches as compared to the HF result, which corresponds to the
experimentally observed order of the outer valence ionization
potentials \cite{JamorskiCasidaSalahub:96}.
As far as the multipole moments are concerned, the largest discrepancy
between the x-only KLI and HF approaches occurs for the total
hexadecapole moment of N$_{2}$, where the results differ by 8.8\%. 
In this case, the Slater approximation gives a value differing only by
3.7\% from the HF one. The $1/r$ expectation values obtained with
the HF and x-only KLI method are almost
identical, differing by only a few hundreths of a precent with the
exception of the one for the Hydrogen nucleus in FH, where the
difference is an order of magnitude larger. In this case, both the
Slater as well as the x-only LDA approximations give values closer to
the HF results.

The Slater method gives - with a few exceptions mentioned above -
values for E$_{\rm TOT}$, $\varepsilon_{\rm 
HOMO}$,the multipole moments and $1/r$ expectation values which differ
to a larger extent from 
both the KLI and HF results than the latter from each other. From the
energy eigenvalues of the inner orbitals, on the other hand, it is
obvious that the Slater  
exchange potential $V_{{\rm x}\sigma}^{\rm S} \br$ is deeper than the
one obtained in the KLI method, giving results closer the HF ones.

Finally, the x-only LDA results differ more strongly from the other
methods, yielding much higher total energies. The difference is most
pronounced for the  values of $\varepsilon_{\rm HOMO}$, which are
roughly twice 
as large as the ones from any of the other methods. This is due to the
wrong exponential decay of $V_{{\rm x}\sigma}^{\rm LDA} \br$ for
large $r$. 

We point out that the bulk part of the difference between
the x-only KLI and the HF results are not due to the KLI
approximation, but to the different nature of the HF and the DFT
approaches. This is an established fact for 
atomic systems 
\cite{KriegerLiIafrate:92,KriegerLiIafrate:92a, LiKriegerIafrate:93} 
and we see no reason why
it should not hold for molecular systems as well.
We mention that the difference between the HF and the exact x-only DFT
results also implies that the {\em exact\/} quantum chemical correlation
energy and the {\em exact\/}  DFT correlation energy are not identical
\cite{GrossPetersilkaGrabo:96}.

\section{\sf  Correlation contributions to the OEP} \label{Correlation
contributions to the OEP} 

\setlength{\textfloatsep}{0pt}
\renewcommand{\textfraction}{0.01}
\renewcommand{\topfraction}{0.99}

The inclusion of correlation effects into the OEP scheme is
straightforward, as anticipated by the indices ${\rm xc}$ in section
\ref{Basic formalism}, once an explicit functional for 
$E_{c}\left[\{\varphi_{j \tau} \} \right]$ 
has been specified. It has been shown
\cite{GraboGross:95, GrossPetersilkaGrabo:96} that the
orbital-dependent Colle-Salvetti functional \cite{ColleSalvetti:75,
ColleSalvetti:79} is well
suited for this purpose.
It yields excellent results for atoms,
surpassing the accurracy of conventional Kohn-Sham calculations. In
this approximation, $E_{c}$ is given by \cite{GraboGross:95}
\bea \label{csec2}
E_{c} \left[\{\varphi_{j \tau} \} \right]
& = &  -  ab \int d^3 r \ \gamma \br \xi \br \Biggl[ \sum_{\sigma}
\rho_{\sigma} \br \sum_{i} \mid \!{\bf \nabla} \varphi_{i\sigma} \br
\! \mid^{2}  
\ - \ \ \frac{1}{4}  \mid \! {\bf \nabla} \rho \br \! \mid^{2} \nonumber \\
&&  \qquad \qquad \qquad \qquad \ - \ \frac{1}{4}  \sum_{\sigma}
\rho_{\sigma}\br \triangle \rho_{\sigma} \br \ 
+ \ \frac{1}{4}  \rho \br \triangle \rho \br \Biggr]  \nonumber \\
&& -a \int d^3 r \ \gamma \br \frac{ \rho \br }{\eta \br} , 
\eea
where
\begin{eqnarray}
\label{gam}
\gamma \br& = & 4 \ \frac{\rho_{\uparrow}\br
\rho_{\downarrow}\br}{\rho \br^{2}}, \\ 
\label{eta}
\eta \br & = & 1 + d \rho \br^{-\frac{1}{3}},\\
\label{xsi}
\xi \br & = & \frac{\rho \br^{-\frac{5}{3}} e^{-c \rho
\br^{-\frac{1}{3}}}}{\eta \br} 
\end{eqnarray}
and
\bed
\begin{array}{ll}
a = 0.04918, \qquad & b = 0.132, \\
c = 0.2533, & d = 0.349 .
\end{array}
\eed

\subsection{\sf  Two-Electron Systems}

\begin{table}[tbh]
\begin{center}
\begin{tabular}{|l|r|r|r|r|}
\hline
 & KLICS & BLYP & PW91 & exact  \\
\hline
H$^{-}$ & 0.5189 & & & 0.5278 \\
He & 2.9033 & 2.9071 & 2.9000 & 2.9037  \\
Li$^{+}$ & 7.2803 & 7.2794 & 7.2676 & 7.2799\\
Be$^{2+}$ & 13.6556 & 13.6500 & 13.6340 & 13.6556 \\
B$^{3+}$ & 22.0301 & 22.0200 & 21.9996 & 22.0310 \\ 
C$^{4+}$ & 32.4045 & 32.3896 & 32.3649 & 32.4062 \\ 
N$^{5+}$ & 44.7788 & 44.7592 & 44.7299 & 44.7814 \\ 
O$^{6+}$ & 59.1531 & 59.1286 & 59.0948 & 59.1566 \\ 
F$^{7+}$ & 75.5274 & 75.4981 & 75.4595 & 75.5317 \\ 
Ne$^{8+}$ & 93.9017 & 93.8675 & 93.8241 & 93.9068 \\
Na$^{9+}$ & 114.2761 & 114.2369 & 114.1886 & 114.2819 \\ 
Mg$^{10+}$ & 136.6505 & 136.6064 & 136.5531 & 136.6569 \\ 
Al$^{11+}$ & 161.0250 & 160.9758 & 160.9175 & 161.0320 \\ 
Si$^{12+}$ & 187.3995 & 187.3453 & 187.2819 & 187.4070 \\ 
P$^{13+}$ & 215.7740 & 215.7147 & 215.6462 & 215.7821 \\
S$^{14+}$ & 246.1485 & 246.0842 & 246.0105 & 246.1571 \\ 
Cl$^{15+}$ & 278.5231 & 278.4536 & 278.3748 & 278.5322 \\
Ar$^{16+}$ & 312.8977 & 312.8231 & 312.7390 & 312.9072 \\
K$^{17+}$ & 349.2723 & 349.1926 & 349.1032 & 349.2822 \\ 
Ca$^{18+}$ & 387.6470 & 387.5620 & 387.4674 & 387.6572 \\
\hline
$\triangle$ & 0.0053 & 0.0450 & 0.0943 & \\
\hline
\end{tabular}
\end{center}
\caption{  Total absolute ground-state energies for the
Helium isoelectronic series
from various self-consistent calculations. $\triangle$ denotes the
mean absolute deviation from the exact values from
\protect\cite{DavidsonEtAl:91} . All 
numbers in atomic units. } 
\label{tab1}  
\vspace{0.5cm} 
\end{table}

In order to study the correlation contributions more thoroughly, we
first concentrate on two-electron atoms for two
reasons. First of all, as pointed out above, the solution of the full
OEP integral equation
for these systems is
identical to the one obtained from the KLI-scheme. As the exchange energy
functional is also known exactly, c.f. equation (\ref{exhf}), the only
error made is due to the approximation for $E_{c}$. Secondly there exist
practically exact solutions \cite{UmrigarGonze:94}
of the two-particle Schr\"odinger equation so that
various DFT-related quantities of interest
can be compared with exact results.

In Table \ref{tab1} we show the total absolute ground-state energies
of the atoms isoelectronic with helium. The first column, denoted by
KLICS, displays the results obtained with the above described method,
including the Colle-Salvetti functional for $E_{c}$ into the OEP
scheme. The next two columns show results obtained with the
conventional Kohn-Sham method for comparison. BLYP denotes the use of
the exchange-energy functional by Becke \cite{Becke:88}
combined with the
correlation energy functional by Lee, Yang and Parr
\cite{LeeYangParr:88}, whereas the third 
column headed PW91 refers to the generalized gradient approximation by
Perdew and Wang \cite{PerdewBurkeWang:96}. The exact nonrelativistic 
results in the last column are taken from 
\cite{DavidsonEtAl:91}. Note that there is no convergence for
neagtive ions in the 
conventional Kohn Sham method.
All of our calculations have been performed with a basis-set-free, 
fully numerical atomic
code which solves the radial Schr\"odinger equation (\ref{1p-eq}) by
the Numerov-method as described in \cite{Froese-Fischer:77}. The angular
parts are treated analytically.

\begin{figure}
\begin{picture}(15,9)
\put(-6,3.5){\makebox(15,9){
\includegraphics{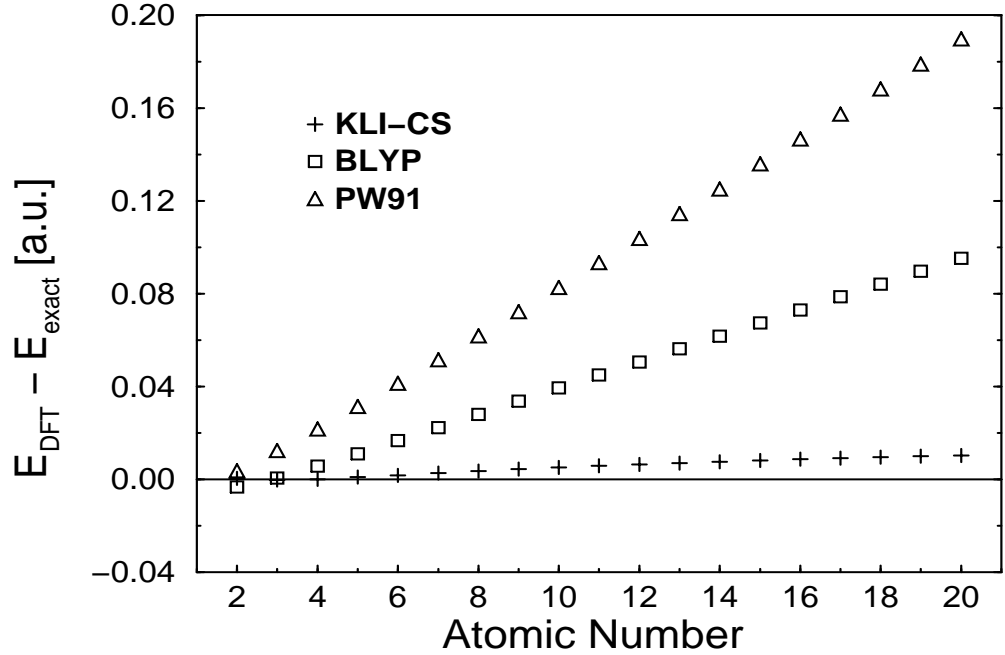} }}
\end{picture}
\caption{Energy differences corresponding to Table \protect\ref{tab1}} 
\label{TabHeIsoErr}
\vspace{0.5cm}
\end{figure}

In Figure \ref{TabHeIsoErr}, we have plotted the errors $E_{{\rm
tot}}^{\rm DFT} - 
E_{{\rm tot}}^{\rm exact}$ of the numbers shown in Table \ref{tab1}. It is
obvious that the KLICS scheme gives superior results. The mean
absolute error, denoted by $\triangle$, is smaller by an order of
magnitude for the KLICS results as compared to the two conventional Kohn
Sham approaches. This comes as no surprise, as the exchange
part is treated exactly in the OEP method, whereas only approximative
functionals can be used in the Kohn-Shame scheme. 
As may be read off
 Table \ref{tabxc}, where we show the exchange and correlation
contributions to the total energy seperately for systems where
exact values are available, an error cancellation occurs in the
BLYP and PW91 approaches - the exchange energies being too large and the
correlation energies being to small in magnitude. The KLICS results
for these two quantities are clearly much better. 
However, for the highly charged two-electron ions the quality of the
results decreases substantially in all approaches. For these systems
the LYP-functional appears to perform best. 

\begin{table}[htbp]
\begin{center}
\begin{tabular}{|l|l|r|r|r|r|}
\hline
 & element & KLICS & BLYP & PW91 & exact \\
\hline
& H$^{-}$ & 0.4053 & & & 0.3809 \\
& He & 1.0275 & 1.0183 & 1.0095 & 1.0246 \\ 
$-E_{x}$ & Be$^{2+}$ & 2.2768 & 2.2573 & 2.2367 & 2.2766 \\
& Ne$^{8+}$ & 6.0272  & 5.9749 & 5.9189 & 6.0275 \\
& Hg$^{78+}$ & 49.7779 & 49.3412 & 48.8806 & 49.7779 \\
\hline
\hline
& H$^{-}$ & 0.0312 & & & 0.0420 \\
& He & 0.0416 & 0.0437 & 0.0450 & 0.0421 \\
$-E_{c}$& Be$^{2+}$ & 0.0442 & 0.0493 & 0.0530 & 0.0443 \\
& Ne$^{8+}$ & 0.0406 & 0.0504 & 0.0615 & 0.0457 \\
& Hg$^{78+}$ & 0.0276 & 0.0506 & 0.0805 & 0.0465 \\ 
\hline
\end{tabular}
\end{center}
\caption{  Exchange and correlation energies from various
approximations. Exact values from \protect\cite{UmrigarGonze:94}. All
values in atomic units.} 
\label{tabxc}
\end{table}

In order to assess the quality of the xc potentials resulting from
various approximate functionals, it is informative to look at the
highest occupied orbital energy of the system. In an {\em exact\/}
implementation of DFT this value should be equal to the ionization
potential of the system. Therefore, the resulting values from {\em
approximate\/} schemes are an indication of the quality of the corresponding
xc potential. From Table \ref{tabheiso}, where we have listed these
numbers for various self-consistent approximations together with the
exact ones, it is obvious that the KLICS scheme performs much better
than the conventional Kohn-Sham schemes. 
The difference is less pronounced for the highly charged ions as
the nuclear potential becomes the dominant term in the Kohn-Sham
potential (\ref{vtot}). 
A glance at the second column,
in which we give the corresponding values from an {\em x-only\/} KLI
calculation, shows, however, that the reason for the superior quality
is due to the inclusion of the exact exchange in the KLI scheme, which
results in the correct $-1/r$ asymptotic behaviour of the xc
potential. In fact, adding the Colle-Salvetti formula for the
correlation energy slightly worsens the results, as may be seen by comparing
the  
second and third columns: The correlation contribution lowers
the already too small values from the x-only calculations for the highest
occupied orbital energy even more. 

\begin{table}[htbp]
\begin{center}
\begin{tabular}{|l|r|r|r|r|r|}
\hline
 & KLI & KLICS  &  BLYP  &  PW91  & exact \\ 
& x-only & xc & xc & xc & \\
\hline
 He        &   0.9180 &   0.9446 &   0.5849 &   0.5833 & 0.9037  \\
 Li$^{+}$  &   2.7924 &   2.8227 &   2.2312 &   2.2269 & 2.7799  \\
 Be$^{2+}$ &   5.6671 &   5.6992 &   4.8760 &   4.8701 & 5.6556  \\
 B$^{3+}$  &   9.5420 &   9.5751 &   8.5201 &   8.5129 & 9.5310  \\
 C$^{4+}$  &  14.4169 &  14.4507 &  13.1638 &  13.1554 & 14.4062  \\
 N$^{5+}$  &  20.2918 &  20.3261 &  18.8072 &  18.7978 & 20.2814  \\
 O$^{6+}$  &  27.1668 &  27.2014 &  25.4504 &  25.4401 & 27.1566  \\
 F$^{7+}$  &  35.0418 &  35.0766 &  33.0935 &  33.0823 & 35.0317  \\
 Ne$^{8+}$ &  43.9167 &  43.9517 &  41.7366 &  41.7245 & 43.9068  \\
 Na$^{9+}$ &  53.7917 &  53.8269 &  51.3796 &  51.3666 & 53.7819  \\
 Mg$^{10+}$&  64.6667 &  64.7020 &  62.0225 &  62.0086 & 64.6569  \\
 Al$^{11+}$&  76.5417 &  76.5770 &  73.6654 &  73.6506 & 76.5320  \\
 Si$^{12+}$&  89.4167 &  89.4521 &  86.3083 &  86.2926 & 89.4071  \\
 P$^{13+}$ & 103.2917 & 103.3272 &  99.9511 &  99.9345 & 103.2821  \\
 S$^{14+}$ & 118.1666 & 118.2022 & 114.5939 & 114.5764 & 118.1571  \\
 Cl$^{15+}$& 134.0416 & 134.0773 & 130.2367 & 130.2183 & 134.0322  \\
 Ar$^{16+}$& 150.9166 & 150.9523 & 146.8795 & 146.8602 & 150.9072  \\
 K$^{17+}$ & 168.7916 & 168.8273 & 164.5223 & 164.5021 & 168.7822  \\
 Ca$^{18+}$& 187.6666 & 187.7024 & 183.1650 & 183.1439 & 187.6572  \\
\hline
\end{tabular}
\end{center}
\caption{  Absolute highest occupied orbital energies from various 
self-consistent calculations. Exact values  calculated from 
\protect\cite{DavidsonEtAl:91}. All 
values in atomic units.} 
\label{tabheiso} 
\end{table}

\begin{figure}
\begin{picture}(15,9)
\put(-6,3.5){\makebox(15,9){
\includegraphics{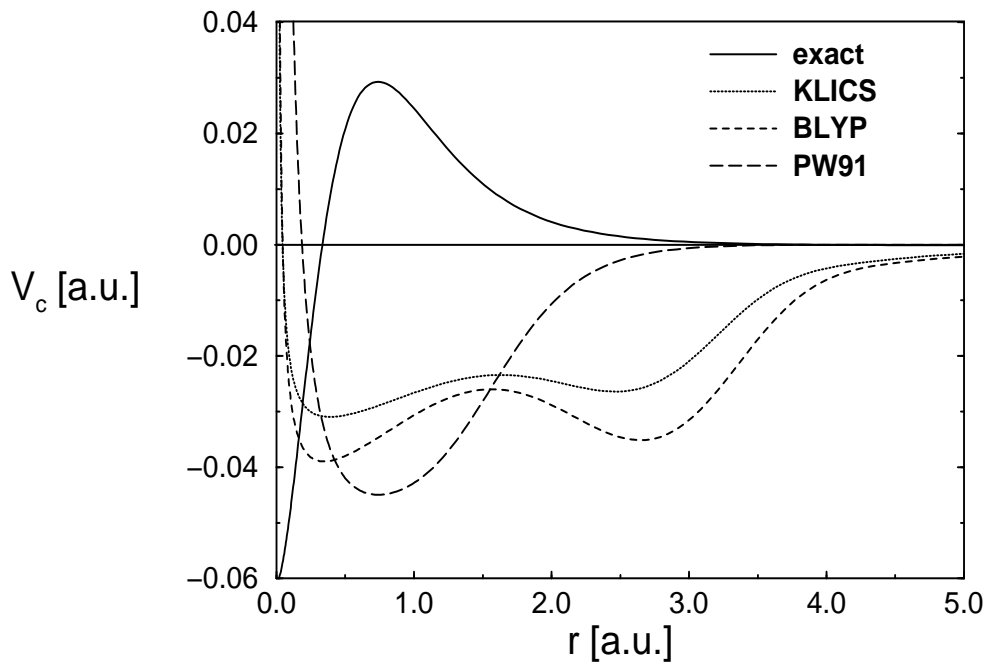} }}
\end{picture}
\caption{ Comparison of the exact and self-consistently calculated
correlation potentials of helium} 
\label{figvc}
\vspace{0.5cm}
\end{figure}

The error in the correlation potential responsible for this behaviour
is clearly visible from Figure
\ref{figvc} where we plot the exact \cite{UmrigarGonze:94}
and various self-consistent correlation
potentials. The potential obtained with the Colle-Salvetti functional
within the KLICS scheme shows the same deficiencies as the conventional
density functionals: Instead of a maximum with positive values of the
potential, the approximations possess one or even two minima and spurious
divergences occur at the origin, which may be traced back to 
gradients of the density and of the one-particle orbitals occuring in
the various correlation energy functionals.
The need for further improvement of the correlation energy functional
in this respect is obvious.

\subsection{\sf  Beryllium and Neon Isoelectronic Series}

\begin{table}[bth]
\begin{center}
\begin{tabular}{|l|r|r|r|r|}
\hline
 & KLICS &  BLYP  &  PW91  & exact \\ 
\hline
     Be              &    14.6651 &    14.6615 &    14.6479 & 14.6674 \\
     B$^{+}$         &    24.3427 &    24.3366 &    24.3160 & 24.3489 \\
     C$^{2+}$        &    36.5224 &    36.5143 &    36.4881 & 36.5349 \\
     N$^{3+}$        &    51.2025 &    51.1927 &    51.1618 & 51.2228 \\
     O$^{4+}$        &    68.3825 &    68.3713 &    68.3362 & 68.4117 \\
     F$^{5+}$        &    88.0624 &    88.0499 &    88.0110 & 88.1011 \\
     Ne$^{6+}$       &   110.2420 &   110.2285 &   110.1859 & 110.2909 \\
     Na$^{7+}$       &   134.9216 &   134.9071 &   134.8610 & 134.9809 \\
     Mg$^{8+}$       &   162.1010 &   162.0857 &   162.0361 & 162.1710 \\
     Al$^{9+}$       &   191.7803 &   191.7642 &   191.7113 & 191.8613 \\
     Si$^{10+}$      &   223.9595 &   223.9427 &   223.8864 & 224.0516 \\
     P$^{11+}$       &   258.6387 &   258.6212 &   258.5616 & 258.7420 \\
     S$^{12+}$       &   295.8178 &   295.7996 &   295.7367 & 295.9324 \\
     Cl$^{13+}$      &   335.4968 &   335.4781 &   335.4119 & 335.6229 \\
     Ar$^{14+}$      &   377.6758 &   377.6566 &   377.5870 & 377.8134 \\
     K$^{15+}$       &   422.3548 &   422.3350 &   422.2621 & 422.5040 \\
     Ca$^{16+}$      &   469.5338 &   469.5134 &   469.4372 & 469.6946 \\
     Sc$^{17+}$      &   519.2127 &   519.1919 &   519.1122 & 519.3851 \\
     Ti$^{18+}$      &   571.3917 &   571.3703 &   571.2873 & 571.5757 \\
     V$^{19+}$       &   626.0706 &   626.0487 &   625.9623 & 626.2663 \\
     Cr$^{20+}$      &   683.2495 &   683.2271 &   683.1373 & 683.4570 \\
     Mn$^{21+}$      &   742.9284 &   742.9056 &   742.8123 & 743.1476 \\
     Fe$^{22+}$      &   805.1072 &   805.0840 &   804.9873 & 805.3382 \\
     Co$^{23+}$      &   869.7861 &   869.7624 &   869.6623 & 870.0289 \\
     Ni$^{24+}$      &   936.9650 &   936.9408 &   936.8373 & 937.2195 \\
\hline
 $\triangle$ &   0.1180 &   0.1352 &   0.1973 &      \\
\hline
\end{tabular}
\end{center}
\caption{  Total absolute ground-state energies for the
Beryllium isoelectronic series
from various self-consistent calculations. 
$\triangle$ denotes the
mean absolute deviation from the exact values from
\protect\cite{ChakravortyEtAl:93} . All 
numbers in atomic units. }  
\label{TabBeIso}
\vspace{0.5cm}
\end{table}

\begin{table}[bth]
\begin{center}
\begin{tabular}{|l|r|r|r|r|}
\hline
 & KLICS  &  BLYP  &  PW91  & exact \\ 
\hline
     Ne              &   128.9202 &   128.9730 &   128.9466 & 128.9376 \\
     Na$^{+}$        &   162.0645 &   162.0956 &   162.0668 & 162.0659 \\
     Mg$^{2+}$       &   199.2291 &   199.2448 &   199.2136 & 199.2204 \\
     Al$^{3+}$       &   240.4071 &   240.4102 &   240.3768 & 240.3914 \\
     Si$^{4}$+       &   285.5945 &   285.5867 &   285.5509 & 285.5738 \\
     P$^{5+}$        &   334.7888 &   334.7712 &   334.7331 & 334.7642 \\
     S$^{6+}$        &   387.9885 &   387.9616 &   387.9212 & 387.9608 \\
     Cl$^{7+}$       &   445.1922 &   445.1567 &   445.1138 & 445.1622 \\
     Ar$^{8+}$       &   506.3993 &   506.3554 &   506.3101 & 506.3673 \\
     K$^{9+}$        &   571.6091 &   571.5570 &   571.5092 & 571.5754 \\
     Ca$^{10+}$      &   640.8211 &   640.7610 &   640.7107 & 640.7861 \\
     Sc$^{11+}$      &   714.0350 &   713.9671 &   713.9141 & 713.9988 \\
     Ti$^{12+}$      &   791.2504 &   791.1748 &   791.1191 & 791.2132 \\
     V$^{13+}$       &   872.4671 &   872.3839 &   872.3255 & 872.4291 \\
     Cr$^{14+}$      &   957.6850 &   957.5942 &   957.5331 & 957.6463 \\
     Mn$^{15+}$      &  1046.9039 &  1046.8056 &  1046.7417 & 1046.8646 \\
     Fe$^{16+}$      &  1140.1237 &  1140.0179 &  1139.9511 & 1140.0838 \\
     Co$^{17+}$      &  1237.3442 &  1237.2309 &  1237.1613 & 1237.3039 \\
     Ni$^{18+}$      &  1338.5654 &  1338.4447 &  1338.3722 & 1338.5247 \\
\hline
$\triangle$ &   0.0293 &   0.0334 &   0.0694 & \\
\hline
\end{tabular}
\end{center}
\caption{  Total absolute ground-state energies for the
Neon isoelectronic series
from various self-consistent calculations. 
$\triangle$ denotes the
mean absolute deviation from the exact values from
\protect\cite{ChakravortyEtAl:93} . All 
numbers in atomic units. }  
\label{TabNeIso}
\end{table}

\begin{figure}[h]
\begin{picture}(15,9)
\put(-6,3.5){\makebox(15,9){
\includegraphics{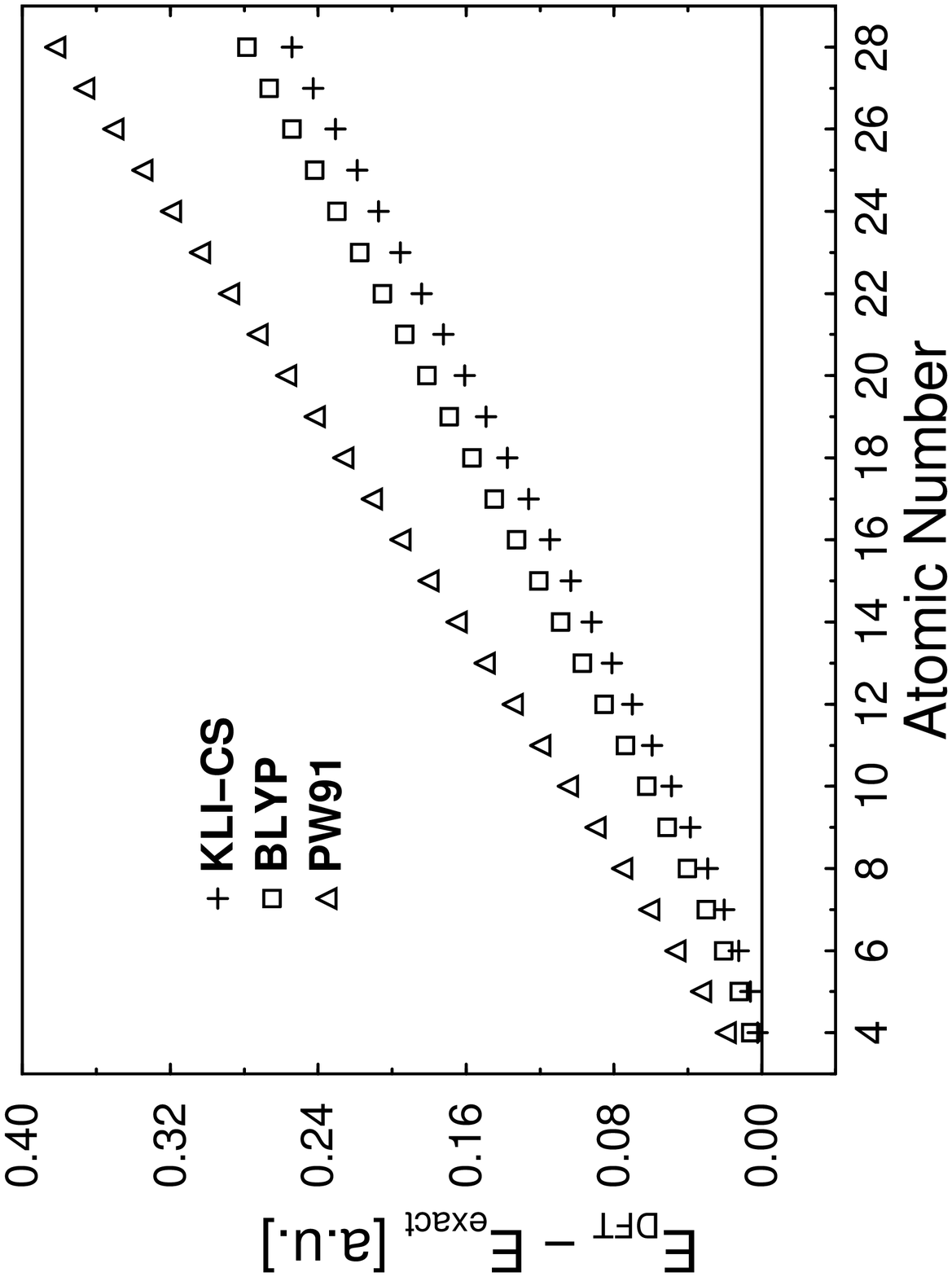} }}
\end{picture}
\caption{ Energy differences corresponding to Table \protect\ref{TabBeIso}. } 
\label{FigBeIsoErr}
\end{figure}

\begin{figure}[h]
\begin{picture}(15,9)
\put(-6,3.5){\makebox(15,9){
\includegraphics{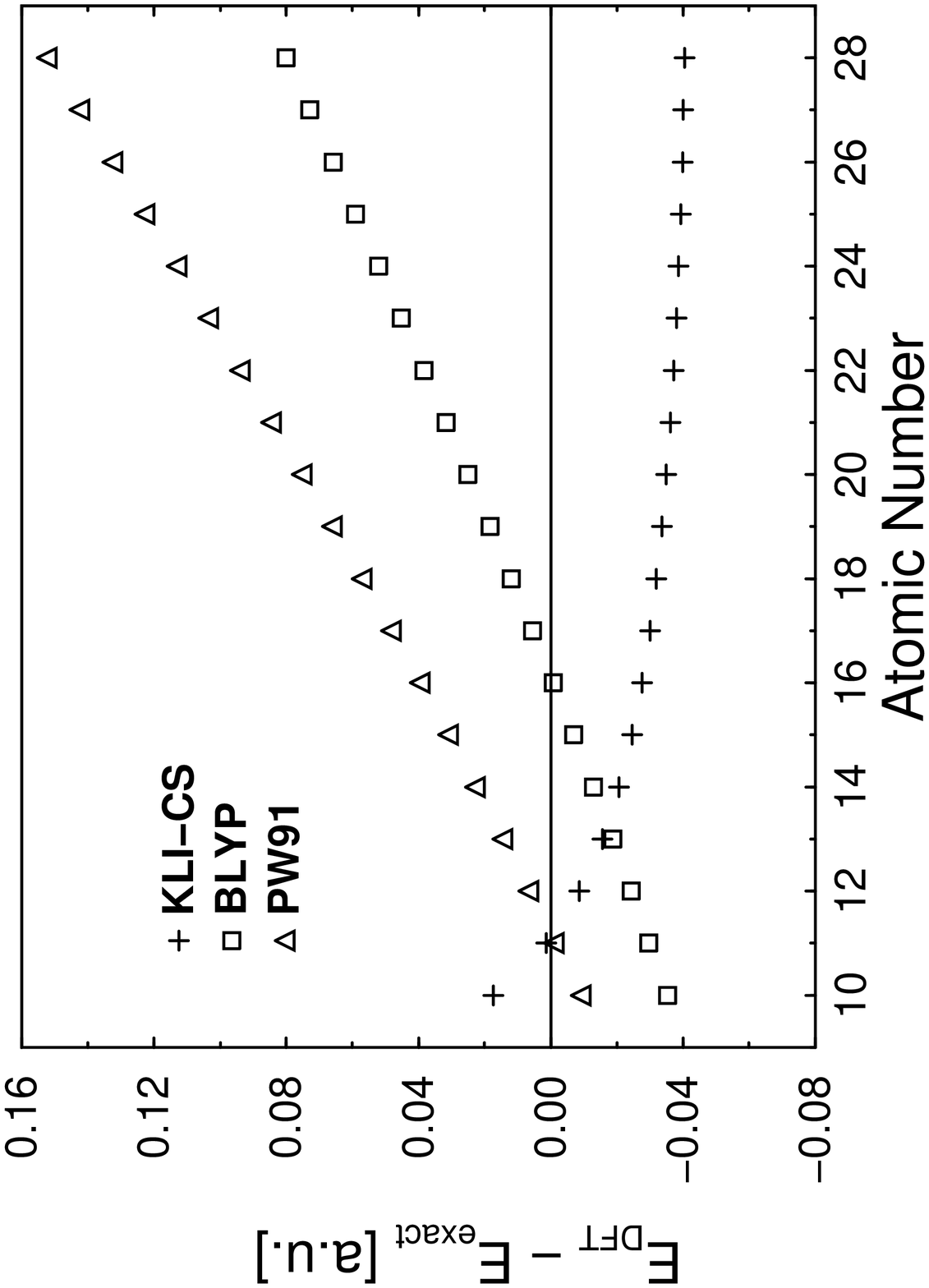} }}
\end{picture}
\caption{ Energy differences corresponding to Table \protect\ref{TabNeIso}} 
\label{FigNeIsoErr}
\end{figure}

For further analysis we have calculated the total ground state
energies of positive ions isoelectronic with Beryllium (shown in Table
\ref{TabBeIso}) and Neon (shown in Table \ref{TabNeIso}). Again, we
compare the various DFT methods with exact data from Ref.
\cite{ChakravortyEtAl:93}.
The errors are plotted in Figures
\ref{FigBeIsoErr} and \ref{FigNeIsoErr}, respectively. The data for both
series show the same trends: The overall best results are obtained with
the KLICS scheme, where the absolute mean deviation $\triangle$ from the
exact values is smallest. The BLYP scheme is only slightly
worse, but the PW91 functional gives errors which are roughly twice as
large as compared to the other DFT approaches. From the plots in Figures
\ref{FigBeIsoErr} and \ref{FigNeIsoErr} it is obvious that these
statements hold for most ions individually. 

There are two other trends to be noted in these results. First we
point out that although the absolute errors rise within the
isoelectronic 
series as the atomic number increases, the percentage errors
remain almost constant. And secondly, the mean absolute error is smaller
by almost an order of magnitude for the ten-electron series compared to
the four-electron series.

The ionization potentials from the various approaches as calculated
from the highest occupied Kohn-Sham orbitals are shown in Tables
\ref{TabBeIsoIon} and \ref{TabNeIsoIon} for the four- and ten-electron
series, respectively. The exact nonrelativistic values have been
calculated from the data given in \cite{ChakravortyEtAl:93}. Due to the 
correct asymptotic 
behaviour of the xc-potential for large $r$ within the OEP scheme it
comes as no surprise that the KLICS data are superior to the
conventional Kohn-Sham approach. The effect of the correlation potential
within the OEP scheme is -- like in the two-electron case -- a lowering
of the energy eigenvalue of the highest occupied orbital, as may be seen
from a comparison of the second and third columns showing the OEP
results in x-only approximation and with inclusion of correlation in the
form of Colle-Salvetti in the KLI-scheme, respectively. As opposed to the
Helium and Neon isoelectronic series, this effect improves the quality
of the results in the Beryllium isoelectronic series. We mention
that the ionization potentials are in much better agreement with the
exact results if they are calculated as ground-state energy differences
\cite{GraboGross:95}.

\begin{table}[bth]
\begin{center}
\begin{tabular}{|l|r|r|r|r|r|}
\hline 
& KLI & KLICS  &  BLYP  &  PW91  & exact \\ 
& x-only & xc & xc & xc & \\
\hline 
Be         &  0.3089 & 0.3294 & 0.2009 & 0.2072 &  0.3426 \\
B$^{+ }$   &  0.8732 & 0.8992 & 0.7129 & 0.7185 &  0.9243 \\
C$^{2+}$   &  1.6933 & 1.7226 & 1.4804 & 1.4856 &  1.7594 \\
N$^{3+}$   &  2.7659 & 2.7975 & 2.5000 & 2.5049 &  2.8459  \\
O$^{4+}$   &  4.0898 & 4.1231 & 3.7706 & 3.7754 &  4.1832 \\
F$^{5+}$   &  5.6644 & 5.6991 & 5.2918 & 5.2964 &  5.7708 \\
Ne$^{6+}$  &  7.4896 & 7.5253 & 7.0633 & 7.0678 &  7.6087 \\
Na$^{7+}$  &  9.5652 & 9.6017 & 9.0850 & 9.0894 &  9.6967 \\
Mg$^{8+ }$ & 11.8910 & 11.9281 & 11.3569 & 11.3611 & 12.0348 \\
Al$^{9+ }$ & 14.4669 & 14.5047 & 13.8788 & 13.8829 & 14.6230 \\
Si$^{10+}$ & 17.2930 & 17.3312 & 16.6508 & 16.6548 & 17.4613 \\
P$^{11+}$  & 20.3692 & 20.4078 & 19.6729 & 19.6768 & 20.5496 \\
S$^{12+}$  & 23.6955 & 23.7345 & 22.9450 & 22.9487 & 23.8880 \\
Cl$^{13+}$ & 27.2718 & 27.3111 & 26.4671 & 26.4707 & 27.4763 \\
Ar$^{14+}$ & 31.0982 & 31.1378 & 30.2393 & 30.2427 & 31.3147 \\
K$^{15+}$  & 35.1747 & 35.2145 & 34.2615 & 34.2647 & 35.4031 \\
Ca$^{16+}$ & 39.5012 & 39.5412 & 38.5336 & 38.5367 & 39.7416 \\
Sc$^{17+}$ & 44.0777 & 44.1179 & 43.0558 & 43.0587 & 44.3300 \\
Ti$^{18+}$ & 48.9043 & 48.9447 & 47.8280 & 47.8307 & 49.1684 \\
V$^{19+}$  & 53.9808 & 54.0214 & 52.8502 & 52.8527 & 54.2569 \\
Cr$^{20+}$ & 59.3074 & 59.3481 & 58.1224 & 58.1247 & 59.5954 \\
Mn$^{21+}$ & 64.8840 & 64.9249 & 63.6447 & 63.6467 & 65.1838 \\
Fe$^{22+}$ & 70.7107 & 70.7516 & 69.4169 & 69.4187 & 71.0223 \\
Co$^{23}$+ & 76.7873 & 76.8284 & 75.4391 & 75.4408 & 77.1108 \\
Ni$^{24+}$ & 83.1140 & 83.1551 & 81.7113 & 81.7128 & 83.4493 \\
\hline
\end{tabular}
\end{center}
\caption{  Ionization potentials from highest occupied Kohn-Sham
orbital energies for the 
Beryllium isoelectronic series
from various self-consistent calculations. 
Exact nonrelativistic values calculated from 
\protect\cite{ChakravortyEtAl:93} . All 
numbers in atomic units. }  
\label{TabBeIsoIon}
\end{table}

\begin{table}[bth]
\begin{center}
\begin{tabular}{|l|r|r|r|r|r|}
\hline 
& KLI & KLICS  &  BLYP  &  PW91  & exact \\ 
& x-only & xc & xc & xc & \\
\hline 
 Ne          & 0.8494 & 0.8841 & 0.4914 & 0.4942 & 0.7945 \\
 Na$^{+ }$   & 1.7959 & 1.8340 & 1.3377 & 1.3416 & 1.7410 \\
 Mg$^{2+ }$  & 3.0047 & 3.0450 & 2.4531 & 2.4579 & 2.9499 \\
 Al$^{3+ }$  & 4.4706 & 4.5125 & 3.8285 & 3.8339 & 4.4161 \\
 Si$^{4+}$   & 6.1912 & 6.2343 & 5.4601 & 5.4661 & 6.1371 \\
 P$^{5+ }$   & 8.1651 & 8.2091 & 7.3458 & 7.3524 & 8.1112 \\
 S$^{6+}$    & 10.3914 & 10.4362 & 9.4846 & 9.4917 & 10.3378  \\
 Cl$^{7+}$   & 12.8696 & 12.9150 & 11.8757 & 11.8832 & 12.8162 \\
 Ar$^{8+}$   & 15.5992 & 15.6451 & 14.5185 & 14.5264 & 15.5460 \\
 K$^{9+ }$   & 18.5800 & 18.6263 & 17.4126 & 17.4209 & 18.5270 \\
 Ca$^{10+}$  & 21.8118 & 21.8585 & 20.5579 & 20.5665 & 21.7589 \\
 Sc$^{11+}$  & 25.2943 & 25.3413 & 23.9541 & 23.9630 & 25.2416 \\
 Ti$^{12+}$  & 29.0274 & 29.0747 & 27.6010 & 27.6102 & 28.9748 \\
 V$^{13+}$   & 33.0112 & 33.0587 & 31.4986 & 31.5080 & 32.9586 \\
 Cr$^{14}$+  & 37.2453 & 37.2931 & 35.6467 & 35.6563 & 37.1929 \\
 Mn$^{15+}$  & 41.7299 & 41.7779 & 40.0453 & 40.0551 & 41.6776 \\
 Fe$^{16}$+  & 46.4649 & 46.5130 & 44.6943 & 44.7042 & 46.4126 \\
 Co$^{17+}$  & 51.4501 & 51.4984 & 49.5936 & 49.6037 & 51.3979 \\
 Ni$^{18+}$  & 56.6856 & 56.7341 & 54.7432 & 54.7535 & 56.6335 \\
\hline
\end{tabular}
\end{center}
\caption{  Ionization potentials from highest occupied Kohn-Sham
orbital energies for the 
Neon isoelectronic series
from various self-consistent calculations. 
Exact nonrelativistic values calculated from 
\protect\cite{ChakravortyEtAl:93} . All 
numbers in atomic units. }  
\label{TabNeIsoIon}
\end{table}

\clearpage

\section{\sf  Conclusions}

Our calculations for molecular systems reveal that the KLI approach is
also feasible for more complex systems and gives results of
similar quality as for atoms. We expect that an inclusion of
correlation effects will result in a highly accurate DFT scheme. The
studies of correlation contributions to atomic systems show that further
improvement of the presently available correlation-energy functionals is 
necessary.

\section{\sf  Acknowledgments}

We would like to thank Dr.~D.~Sundholm and Professor P.~Pyykk\"o for
providing us with their two-dimensional X$\alpha$ code for molecules and
for the warm hospitality during a stay of one of us (T.G.) in
Helsinki. Numerous discussions with Dr.~D.~Sundholm were extremely
valuable. Dr.~E.~Engel supplied us with a conventional 
Kohn-Sham computer code, Professor
C.~Umrigar with the exact densities and Kohn-Sham 
potentials of two-electron atoms and 
Professor J.~Perdew with the PW91 xc subroutine. Many helpful
discussions with M. Petersilka are gratefully acknowledged. 
This work was supported by the Deutsche Forschungsgemeinschaft. 



\end{sf}
\end{document}